\documentstyle[12pt]{article}

\begin{document}

\newcommand{\be}{\begin{equation}}
\newcommand{\ee}{\end{equation}}
\newcommand{\bea}{\begin{eqnarray}}
\newcommand{\eea}{\end{eqnarray}}
\renewcommand{\topfraction}{1.}

\setcounter{page}{0}
\title{{Reduction of internal  
degrees of freedom in the large $N$ limit 
in matrix models}\thanks{Work supported by a 
Spanish M.E.C. Postdoctoral Fellowship}}

\author{{\bf \'Oscar Diego}\thanks{e-mail: diego@nbivms.nbi.dk} \\
        {\em The Niels Bohr Institute } \\
        {\em Blegdamsvej 17, DK-2100 Copenhagen \O, Denmark }}

\date{\mbox{ }}

\maketitle

\thispagestyle{empty}

\begin{abstract}
In this paper the large $N$ limit of one hermitian 
matrix models 
coupled to an external matrix is considered. It is shown 
that in the large $N$ limit the number of degrees of freedom 
are reduced to be order $N$ even though it is 
order $N^{2}$ for finite $N$. 
It is claimed that this result is the origin of the 
factorization of observables in the path integral formalism.
\end{abstract}

\begin{flushright}
\vspace{-14.5 cm} {NBI-HE-97-23}
\end{flushright}
\baselineskip=21pt
\vfill
\newpage
\setcounter{page}{1}

{\bf 1. Introduction} 

It is a well known fact that in the large $N$ limit the 
vacuum expectation value of a finite product of 
observables is the product of vacuum expectation values. 
It is very easy, almost trivial, to prove this 
remarkable factorization in the Feynman diagram 
approach\cite{thoof}. It is also possible to prove 
it in the canonical formalism\cite{yaff}. But in the 
path integral formalism it is not clear how this 
factorization holds. This factorization suggests that 
only one configuration enter into the path integral. 
This is true for vector models but in matrix 
models the number of degrees of freedom is order $N^{2}$ 
and the expansion parameter of the large $N$ limit is 
$1/N^{2}$, therefore the quantum corrections 
to any classical configuration must be order one and 
fluctuations to all orders must enter into the path integral. 
This explains why is so difficult 
to find the master field: because it is not given by 
one configuration but arises from 
several non equivalent configurations. For instance, in one 
hermitian matrix 
models the partition function is given by 
\be
Z = \int D \Phi \exp \{ -N tr [V(\Phi)] \}.
\ee
The potential is a function of $N$ degrees of freedom only: the 
eigenvalues of $\Phi$. Actually
\be
Z = \int D U D \Lambda  \Delta^{2} \exp \{ -N tr [V(\Lambda)] \}
\ee
where $\Lambda$ is the diagonal matrix of the eigenvalues of $\Phi$, 
$\Delta$ is a Van der Monde determinant and the integration over 
the set of unitary matrices $D U$ is trivial. The 
leading contribution to the partition function is given 
by only one configuration of eigenvalues, the master field, 
which is a solution
of the saddle-point condition 
\be
\frac{\partial}{\partial \Lambda_{ii}} 
\left ( \ln \{ \Delta^{2} \exp \{ -N tr [V(\lambda)] \} 
\} \right ) = 0 
\ee
In this case the factorization of observables holds trivially. 

But if we add to the matrix potential a coupling with an external
matrix $A$ then the potential becomes a function of $N^{2}$ 
degrees of freedom. Actually 
\be
Z = \int D U D \Lambda  \Delta^{2} \exp \{ -N tr [V(\Lambda) 
+ U^{\dagger} \Lambda U A ] \}
\label{eq:01}
\ee
We can prove the factorization 
of a product of observables which are functions of the  
eigenvalues, like $< Tr \Phi Tr \Phi >$. 
After the exact integration over $D U$ the partition 
function becomes 
\be
Z ( A ) \propto \int D \Lambda \Delta \exp \{ -N tr [V(\Lambda) 
+ \Lambda A ] \}
\ee
Only one configuration 
of eigenvalues yields the leading contribution to the 
partition function in the large $N$ limit. 
Therefore we can also define a master 
field in this case but first we must perform the 
integration over $N^{2}$ degrees of freedom exactly, in 
other words all configurations of the 
unitary matrix variable $U$ enter into the master field.
  
The existence of a master field for the eigenvalues explains the 
factorization of observables like $< Tr \Phi Tr \Phi>$. 
But for observables like 
\be
< Tr ( \Phi A ) Tr ( \Phi A) > = < Tr ( U^{\dagger} \Lambda U A ) 
Tr ( U^{\dagger} \Lambda U A ) > 
\ee
the factorization is difficult to understand. Of course we know 
that the factorization holds, we can prove it perturbatively, 
but we do not understand the factorization of observables 
nonperturbatively in the path integral approach. 
We can expect that in the large $N$ limit 
the partition function is given by the integral
representation of finite $N$ but restricted over 
some subset of configurations. For instance (\ref{eq:01}) 
in the large $N$ limit should be    
\be
Z = \int_{\Gamma_{\Lambda}} D \lambda 
\int_{\Gamma_{U}}  D U   \Delta^{2} \exp \{ -N tr [V(\Lambda) 
+ U^{\dagger} \Lambda U A ] \}
\ee
where $\Gamma_{\lambda}$ and $\Gamma_{U}$ are 
subsets of the set of all eigenvalues and of all 
unitary matrices respectively. 
We know that there is only one configuration in 
the subset $\Gamma_{\lambda}$. The factorization 
\be
< Tr \Phi A Tr \Phi A > = < Tr \Phi A > < Tr \Phi A > 
\label{eq:fac}
\ee
suggests that there is only one configuration in $\Gamma_{U}$. 
But at first sight this is not possible because the 
set of unitary matrices is labeled by $N^{2}$ 
degrees of freedom. But if there are several 
nonequivalent configurations in $\Gamma_{U}$ 
the 
factorization of observables like (\ref{eq:fac}) 
becomes very unnatural.   
This old problem has been pointed out by 
Itzykson and Zuber in\cite{Itz} several years ago. 

In order to understand this problem we must 
find the configurations which enter into the path 
integral in the large $N$ limit. In this paper we 
will study one hermitian matrix models coupled 
to an external matrix. Actually we will study:
\bea
Z & = & \int D \Phi \exp \left \{ -N Tr [ V (\Phi) + 
A \Phi ] \right \}
\nonumber \\
Z & = & \int D \Phi \exp \left \{ -N Tr [ V (\Phi) + 
A \Phi A \Phi ] 
\right \}
\eea
The first model has been solved in\cite{Itz,Gross} 
and the second in\cite{Kaz}. But we can not read from the 
solutions of the above models 
the configurations that yield the leading 
contribution to the large $N$ limit. In the usual approaches 
one must perform an integration over a subset of variables 
exactly.  In order to understand the factorization of observables 
we must study the large $N$ limit before performing any 
integration. 

We will show that, in the above matrix models, 
the configurations which enter 
into the path integral in the large $N$ limit 
can be labeled with only $N$ 
variables. For instance the subset $\Gamma_{U}$ is 
a subset of dimension $N$. Therefore the quantum corrections to 
any classical configuration are subleading in the 
large $N$ limit. Hence only one configuration yields 
the leading contribution to the 
path integral in the large $N$ limit and the 
factorization of observables follows trivially.

{\bf 2. Integration over the unitary matrix 
in the large N limit} 

Let us start with the following matrix model 
\be
Z(A) = \int D \Phi \exp \left \{ - N [ Tr( \Phi A ) 
+ Tr V (\Phi) ] 
\right \}.
\label{eq:10}
\ee
Without lost of generality we can consider diagonal 
external matrices.

Let us perform the change of variables 
\be
\Phi \rightarrow U^{\dagger} \Lambda U,
\ee
where $\Lambda$ is the diagonal matrix of the 
eigenvalues of $\Phi$ and $U$ is a unitary 
matrix. The partition function becomes:
\be
Z(A) = \int D \Lambda \Delta^{2} 
\exp \{ - N Tr V(\Lambda) \} \tilde{Z} 
(A,\Lambda),
\label{eq:11}
\ee
where 
\be
\tilde{Z}(A, \Lambda) = \int D U \exp 
\{ - N Tr ( U^{\dagger} \Lambda U A ) \}. 
\label{eq:12}
\ee
This integral can be solved with the Itzykson-Zuber 
formula\cite{Itz}. But we will show that in the 
large $N$ limit the above integral with 
$\Lambda$ fixed can be reduced to an integral 
over unitary matrices with only order $N$ 
nonvanishing elements. Hence, as in vector models, 
we can use saddle-point methods to calculate it.     

In the introduction I have claimed that at first sight 
an integral like (\ref{eq:12}) cannot be solved with a saddle-point 
method. This is because any correction must be order the number
of degrees of freedom which, in matrix models, is the 
same as the inverse of the expansion parameter. Therefore 
any correction must be order one. But the above 
argument does not take into account 
the compactness of the unitary set. 
Unitary matrices must satisfy the constraint
\be
U U^{\dagger} = I
\ee
Therefore, the absolute values of the 
elements of an arbitrary unitary matrix
must decrease when $N$ increase. Hence
the range of variation of $U$ decrease with $N$. 
But there are exceptions 
to this rule. For instance the absolute values of the 
non zero elements of a diagonal matrix are order one 
no matter how bigger is $N$. Actually for bidiagonal, 
tridiagonal and in general matrices with a finite 
number of nonzero elements in each row and column, 
the absolute values of its non zero elements are order one.
Therefore we can expect that these configurations must play 
some special role in the large $N$ limit. Actually 
I will show that these configurations yield the leading 
contribution to the partition function in the large $N$ limit. 

There are two kind of unitary matrices. Unitary matrices 
with order $N^{2} - N$ 
elements so smaller that they decoupled from the 
matrix potential in the large $N$ limit, let us 
call them unitary matrices with $N$ degrees of freedom,  
and unitary matrices where order $N^{2}$ elements 
will be coupled between them through 
the matrix potential in the large $N$ limit, let us call them 
unitary matrices with $N^{2}$ degrees of freedom. 
Hence the partition function should be the sum of two terms
\be
Z = Z_{uncoupled} + Z_{coupled}
\ee
where $Z_{uncoupled}$ is the partition function 
restricted over the subset of unitary matrices with $N$ 
degrees of freedom and $Z_{coupled}$ is restricted 
over the subset of unitary matrices with $N^{2}$ degrees of 
freedom. 

It is not possible 
to solve $Z_{coupled}$ but it is trivial that
\be
Z_{coupled} \le Vol \exp \{ - N 
Tr ( U^{\dagger}_{S} \Lambda U_{S} A ) \}
\ee
where $Vol$ is the volume of the 
subset of unitary matrices with $N^{2}$ degrees of 
freedom and $U_{S}$ is the absolute minimum of the 
potential in the subset of 
unitary matrices with $N^{2}$ degrees of freedom. 

The partition function $Z_{uncoupled}$ is given by 
\be
Z_{uncoupled} = A \exp \{ - N^{2} F \}
\ee
where $A$ is the factor which arises after the integration 
over the $N^{2} -N$ degrees of freedom which decoupled from 
the potential and I will show that $F$ 
is the free energy of a matrix model
restricted over unitary matrices with only $N$ nonzero elements. 
I will show that $A$ and $Vol$ are of the same order 
in the large $N$ limit: this is because the compactness 
of the unitary group. Then if 
\be
F < \frac{1}{N} Tr ( U^{\dagger} \Lambda U_{S} A ) 
\ee
then $Z_{uncoupled}$ gives the leading contribution to $Z$. 
And because $Z_{uncoupled}$ can be reduced to an integral 
over $N$ degrees of freedom only; actually I will show that 
\be
Z_{uncoupled} = \int D U \exp 
\{ - N Tr ( U^{\dagger} \Lambda U A ) \}      
\ee
where $D U$ is the measure over unitary matrices with 
only $N$ nonzero elements; there is a master field 
in the variables $U$ and corrections to this 
master field will be subleading in the large $N$ limit. 

In order to classify the unitary matrices 
according to the absolute values of its elements 
we must transform the measure over the holomorphic 
elements 
\be
D U = \prod_{ab} d U_{ab}
\ee
into a measure which includes the absolute values of the 
elements
\be
D U = \prod_{a \le b} d | U_{ab} | \cdots 
\ee

The invariant measure $D U$ can be extended over 
the set of all matrices as follows:
\be
\int D U = \int \prod_{a,b=1}^{N} \left \{
d U_{ab} d U^{*}_{ab}  \delta \left ( \sum_{c=1}^{N} 
U_{ac}U^{*}_{bc} - \delta_{ab} \right ) \right \}. 
\label{eq:22}
\ee
This measure is invariant under the unitary 
transformation. 

Let us perform the following change of variables:
\be
\left ( U_{ab}, U^{*}_{ab} \right ) 
\rightarrow \left ( |U_{ab}| , \theta_{ab} \right ),
\label{eq:23}
\ee
where $\theta_{ab}$ are the phases of the 
elements of the unitary matrix. 
The Jacobian of (\ref{eq:23}) is 
\be
\prod_{a,b=1}^{N} |U_{ab}|.
\label{eq:24}
\ee
Therefore the measure (\ref{eq:22}) becomes 
\be
\int D U = \int \prod_{a,b=1}^{N} \left \{
d |U_{ab}|  \hbox{ } d \theta_{ab} \hbox{ } 
|U_{ab}| \hbox{ } 
\delta \left ( \sum_{c=1}^{N} 
|U_{ac}| |U_{bc}| \exp 
( i \theta_{ac} - i \theta_{bc} ) 
- \delta_{ab} \right ) \right \}
\label{eq:25}
\ee
and the partition function (\ref{eq:12}) becomes:
\bea
\tilde{Z} (A,\Lambda) & = &  \int \prod_{a,b=1}^{N} 
\left ( d |U_{ab}| \hbox{ } d \theta_{ab} \right ) 
\prod_{a,b=1}^{N} |U_{ab}| \nonumber \\
 & & \prod_{a,b=1}^{N} \delta \left [ \sum_{c=1}^{N} 
|U_{ac}| |U_{bc}| 
\exp ( i \theta_{ac} - i \theta_{bc} ) - \delta_{ab} 
\right ]  \nonumber \\
 & & \exp \left \{ - N \sum_{a,b=1}^{N} |U_{ab}|^{2}
\Lambda_{bb} A_{aa} \right \}. 
\label{eq:26}
\eea

The variables $|U_{ab}|$ are restricted by the 
delta functions which yield the following constraints:
\be
\sum_{c=1}^{N} |U_{ac}| |U_{bc}| \exp \left 
( i \theta_{ac} - i \theta_{bc}
\right ) = \delta_{ab}. 
\label{eq:27}
\ee

Let us study now the behaviour of the 
measure (\ref{eq:25}) in the large $N$ limit for 
differents subset of unitary matrices. 
For homogeneous configurations: all $|U_{ab}|$ are 
of the same order, the 
unitary constraints (\ref{eq:27}) restrict the 
variables $\{ |U_{ab}| \}$ to be order $1/\sqrt{N}$. 

Let $Z_{H}$ the partition function (\ref{eq:26}) 
restricted over the set of homogeneous configurations. 
Then there is a trivial bound to $Z_{H}$ given by 
\be
Z_{H} \le Vol \exp \left 
\{ - N Tr ( U^{\dagger}_{s} 
\Lambda U_{s} A ) \right \},
\ee
where $U_{s}$ is the absolute minimum 
of $Tr ( U^{\dagger}_{s} \Lambda U_{s} A )$ 
in the subset of homogeneous configuration 
and $Vol$ is the volume of the subset of 
homogeneous configurations. Because the dimension of 
the subset of homogeneous configurations is $N^{2}$ the 
volume must be order 
\be
Vol \propto (\frac{1}{\sqrt{N}})^{N^{2}}
\ee

Let us study now the partition function $Z_{I}$ 
restricted over the following inhomogeneous 
configurations: for each row and column there are 
a finite number of elements with absolute value order one, 
let us call them large elements, and the others are 
order less than $1/\sqrt{N}$, let us call them small. 

Order less than $1/\sqrt{N}$ means that 
the small elements are order 
\be
| U_{ab}^{small} | \propto ( \frac{1}{\sqrt{N}} )^{(1 + \epsilon) }
\label{eq:n25}
\ee
where $\epsilon$ is a positive number which goes to zero in the 
large $N$ limit. 

The are three contribution to the constraints (\ref{eq:27}).
The contribution which comes from the large elements 
\be
\sum_{c} 
|U_{ac}^{large}| |U_{bc}^{large}| \exp \left 
( i \theta_{ac} - i \theta_{bc}
\right ). 
\ee
This is order one because the sum over $c$ can only have a finite 
number of terms. 
Let us remark that these inhomogeneous matrices must have only 
a finite number of large elements in each row a column, otherwise
the above contribution becomes order greater than one for $a = b$ 
and the matrices are not unitary.  
      
The contribution which comes from the 
small elements 
\be
\sum_{c} 
|U_{ac}^{small}| |U_{bc}^{small}| \exp \left 
( i \theta_{ac} - i \theta_{bc}
\right ). 
\ee
This is order 
\be
\left ( \frac{1}{N}
\right )^{\epsilon}
\ee
because now the sums over $c$ runs over order $N$ elements. 
And the contribution which comes from the crossing of large 
and small elements. 
\be
\sum_{c} 
|U_{ac}^{large}| |U_{bc}^{small}| \exp \left 
( i \theta_{ac} - i \theta_{bc}
\right ).
\label{eq:cross} 
\ee
This is order 
\be
\frac{1}{\sqrt{N}}\left ( \frac{1}{\sqrt{N}}
\right )^{\epsilon}
\ee
because the sum over $c$ is over a finite number of terms. 
There are only a finite number of nonzero large 
elements $|U_{ac}^{large}|$ for fixed $a$.

Now let us define $\epsilon$ as 
\be
\epsilon = \frac{1}{\sqrt{\ln{N}}} 
\ee
Then $\epsilon$ goes to zero in the large $N$ limit and 
\be
\left ( \frac{1}{\sqrt{N}} \right )^{\epsilon} \rightarrow 0 
\ee
 
Hence, in the large $N$ limit the unitary constraints 
becomes  
\be
\sum_{c} 
|U_{ac}^{large}| |U_{bc}^{large}| \exp \left 
( i \theta_{ac} - i \theta_{bc}
\right ) = \delta_{ab}. 
\label{eq:consredu}
\ee
Therefore the small elements have decoupled from the 
unitary constraints. Let us remark that the 
large elements can be put on a matrix, 
say $U_{L}$, and 
the constraints (\ref{eq:consredu}) 
are equivalent to the unitarity of $U_{L}$.       

But we cannot perform the large $N$ limit inside 
the Dirac delta function, in other words is not true 
that in the large $N$ limit
\be
\delta ( U U^{\dagger} - I ) \rightarrow 
\delta ( U_{L} U^{\dagger}_{L} - I ) 
\ee
This is because the matrix $U_{L}$, the matrix of the 
large elements only, has only order 
$N$ variables and there are $N^{2}$ delta functions 
in the partition function (\ref{eq:26}).

Let us consider the unitary constraints up to 
the first correction in the large $N$ limit. Then 
\be
\sum_{c} 
|U_{ac}^{large}| |U_{bc}^{large}| \exp \left 
( i \theta_{ac} - i \theta_{bc}
\right ) 
+ 
\sum_{c} 
|U_{ac}^{small}| |U_{bc}^{small}| \exp \left 
( i \theta_{ac} - i \theta_{bc}
\right ) = \delta_{ab}. 
\label{eq:conscor}
\ee
The cross term (\ref{eq:cross}) gives the second correction. Because 
there are only $N$ variables $|U_{ab}^{large}|$, a finite 
number in each row and column, the large variables only appear 
in order $N$ constraints. 
For instance a matrix with $N$ large variables in only one row
gives contribution to all the constraints but such matrices 
are not unitary.   

Therefore, there are $N$ delta 
functions which depend on the large variables and in the large 
$N$ limit they become.
\be
\prod_{(a,b) \in \Gamma }
\delta \left ( \sum_{c} |U_{ac}^{large}| |U_{bc}^{large}| \exp \left 
( i \theta_{ac} - i \theta_{bc}
\right ) - \delta_{ab} \right )
\label{eq:large}
\ee
If we call $I_{N}$ the set given by the first $N$ integers then
$\Gamma$ is a subset of the set $I_{N} \times I_{N}$ and has only 
order $N$ elements. In other words (\ref{eq:large}) is the 
product of order $N$ delta functions. 
  
And there are $N^{2} - N $  delta functions 
which are independent of the large 
variables:
\be
\prod_{(a,b) \in \bar{\Gamma} }
\delta \left ( \sum_{c} |U_{ac}^{small}| |U_{bc}^{small}| \exp \left 
( i \theta_{ac} - i \theta_{bc}
\right ) \right )
\label{eq:small}
\ee
where $\bar{\Gamma}$ is the complementary of $\Gamma$ 
in $I_{N} \times I_{N}$. 
Let us remark that there is a complete decoupling 
between the large and the small variables in the measure. 

Let us define $\delta U$ as the matrix given by the small 
elements only. Then the delta functions (\ref{eq:large}) and 
(\ref{eq:small}) means that 
\bea
(U_{L})_{ac} (U^{\dagger}_{L})_{cb} & = & \delta_{ab} \ \ 
( a , b )  \in \Gamma  \nonumber \\
(\delta U)_{ac} (\delta U^{\dagger})_{cb} & = & 0
\ \ ( a , b ) \in \bar{\Gamma}
\label{eq:delcons}
\eea
     
Let us remark that the second equation in (\ref{eq:delcons}) 
does not mean that $\delta U$ is singular because the indices $a$ 
and $b$ do not run over all integers between one and $N$. 
Actually we can write the second equation as 
\be
\delta U \delta U^{\dagger} = J
\ee
where $J$ is an arbitrary matrix with nonzero elements 
$J_{ab}$ for the labels $ ( a , b ) $ that belongs to $\Gamma$ and 
\be
J_{ab} \propto \left ( \frac{1}{N} \right )^{\epsilon} \ \
( a , b ) \in \Gamma.
\ee

Let us consider the matrix potential now. 
The matrix potential can be split into two parts:
\be
V = - N \sum_{a=1}^{N} \sum_{b \in large } 
|U_{ab}|^{2} \Lambda_{bb} A_{aa}
    - N \sum_{a=1}^{N} \sum_{b \in small } 
|U_{ab}|^{2} \Lambda_{bb} A_{aa}
\label{eq:35}
\ee
The first term is order $N^{2}$ because the sum is 
extended over the variables which are order one 
and the sum over the index $b$ has only a finite 
number of terms. In the second term each sum over 
indices $a$ and $b$ is order $N$, but the order 
of the absolute values are less than $1/\sqrt{N}$ 
and therefore the second term is order less than $N^{2}$. 
Therefore the matrix potential is a function 
of the large variables only in the 
large $N$ limit.

Let us perform the integration over the small 
variables. We must take into account that 
\bea
\delta ( \delta U \delta U^{\dagger} - J ) & = & 
\frac{1}{\det U} \delta ( \delta U^{\dagger} - J (\delta U)^{-1} ) 
\nonumber \\
    					   & = & 
\frac{1}{\sqrt{\det J}} 
\delta ( \delta U^{\dagger} - J (\delta U)^{-1} ) 
\label{eq:n42}
\eea
Hence, taken into account the decoupling of the 
small variables, equations  (\ref{eq:n42}) and  (\ref{eq:n25}), and the 
definition of $\epsilon$, the integration over the small 
variables becomes 
\be
\int D (\delta U) D (\delta U^{\dagger} ) 
\delta ( \delta U \delta U^{\dagger} - J ) \propto 
\left ( \frac{1}{\sqrt{N}} \right )^{ ((N^{2} - N) ( 1 + \epsilon ) 
- \epsilon N )} \rightarrow \left ( \frac{1}{\sqrt{N}} \right )^{N^{2}}
\ee             

Hence after the integration over the small 
variables, $Z_{I}$ becomes in the large $N$ limit
\be
Z_{I} \propto \left ( 
\frac{1}{\sqrt{N}} \right )^{N^{2}}
\int D U_{L} D U^{\dagger}_{L}\delta 
( U^{\dagger}_{L} U_{L} - 1 )  
\exp \left \{ - N Tr ( U_{L}^{\dagger} 
\Lambda U_{L} A ) \right \}
\ee
and after the integration over $D U^{\dagger}_{L}$ 
\be
Z_{I} \propto \left ( 
\frac{1}{\sqrt{N}} \right )^{N^{2}}
\int D U_{L} \exp \left 
\{ - N Tr ( U_{L}^{\dagger} 
\Lambda U_{L} A ) \right \}
\label{eq:36}
\ee
where now $D U_{L}$ is the measure over 
unitary matrices with only order 
$N$ elements nonzero. Let us remark that 
that the number of nonzero elements 
is not fixed but it is order $N$. Therefore the 
dimension of the set of inhomogeneous matrices 
is not a fixed number. This means that 
$D U_{L}$ is not a Riemanian measure in the usual sense. 
This is not important in the large $N$ limit. Actually I will 
show that the integral over $D U_{L}$ is given by only one
configuration in the large $N$ limit. 

But we can also define $D U_{L}$ explicitly as a sum 
\be
D U_{L} =  D_{1} U_{L} + D_{2} U_{L} + \cdots 
\ee
where $D_{1} U_{L}$ is the measure over unitary 
diagonal matrices, $D_{2} U_{L}$ is the measure 
over unitary matrices with two diagonals a so on.
For instance let us consider the set of inhomogeneous 
matrices with large $N$ elements in the main diagonal 
only. Then we perform all the tricks describe above 
and we arrive 
to the partition function restricted over   
inhomogeneous matrices with large elements in the
main diagonal 
\be
Z_{I}^{1} \propto \left ( 
\frac{1}{\sqrt{N}} \right )^{N^{2}}
\int D_{1} U_{L} \exp \left 
\{ - N Tr ( U_{L}^{\dagger} 
\Lambda U_{L} A ) \right \}
\ee
where $D_{1} U_{L}$ is the measure over 
diagonal unitary matrices. In this way we can 
write the partition function $Z_{I}$ as a sum 
over partition function restricted over 
diagonal, bidiagonal and so on. 
The factors in front of each partition function $Z_{I}^{j}$ 
are differents, but all are of the same order in the large 
$N$ limit. These factors are not important because in the 
large $N$ limit only one configurations enter into 
the path integral $Z_{I}$.

The partition function $Z_{I}$ can be 
solved with a saddle-point 
method because depends only on order $N$ 
degrees of freedom. Actually the saddle-point 
equation is given by 
\be
\frac{\partial}{\partial U_{ab}} Tr ( U^{\dagger} \Lambda U A ) = 0  
\label{eq:sp}
\ee
In this case any correction to the saddle-point solutions 
are order the expansion parameter, which is $1/ N^{2}$, times 
the number of degrees of freedom, which are order $N$. Let us remark 
that if we perform any change of variables in (\ref{eq:36}), 
for instance 
\be
(U_{L})_{ab} \rightarrow ( |(U_{L})_{ab}| , \theta_{ab} ), 
\label{eq:ex}
\ee
the saddle-point conditions (\ref{eq:sp}) do not change. 
This is because the Jacobian of any change of variables 
grows as an exponential of $N$, because there are only $N$ degrees
of freedom, for instance the jacobian 
of (\ref{eq:ex}) 
is the product of $N$ elements        
\be
\prod_{(a,b) \in \Gamma} |(U_{L})_{ab}|,
\ee
whereas the potential grows as the exponential
of $N^{2}$. 

Taken into account that $U_{L} U^{\dagger}_{L} = I$, the
saddle point configuration of (\ref{eq:sp}) are given by:
\be
U \Lambda U^{\dagger} A = A U \Lambda U^{\dagger}, 
\label{eq:41}
\ee
or in other words the matrix $\Phi$ must commute 
with the external matrix $A$. Actually (\ref{eq:41}) 
is the condition of commutativity of $\Phi$ and $A$ 
in the basis where the matrix $A$ is diagonal. 

Let us remark that the above equation (\ref{eq:41}) comes
from the fact that $U_{L}$ is a unitary matrix with 
only order $N$ nonzero elements. It is not important 
if all its nonzero elements are order one or $1/\sqrt{N}$. 
Of course a unitary matrix with only order $N$ nonzero 
elements must have $N$ large elements.

The same analysis can be carried out for configurations  
with large elements, small elements and order $1/\sqrt{N}$
elements. For instance configurations 
with order $N$ large elements and order $N^{2} - N$
elements order $1/\sqrt{N}$. Let us remark that there 
is a very important difference between elements 
order $1/\sqrt{N}$ and small 
elements, which are order less than $1/\sqrt{N}$. The former
are coupled with the large elements and between them through the 
potential and the small elements do not enter into the potential 
and do not coupled with the large elements through the 
unitary constraint. Nevertheless we must be 
careful with the above division between large, small 
and homogeneous elements. An element order $1/\sqrt{N}$
enter into the potential only if there are order $N^{2}$
homogeneous elements in the same row or column. 

Therefore
we can construct configurations with order $N^{2}$ 
degrees of freedom coupled between them starting 
with the homogeneous configurations and changing 
a finite number of homogeneous elements in each row an column 
by small elements. In this way we can also construct 
configurations with order $N$ degrees of freedom 
coupled between them starting from the inhomogeneous 
configurations and changing a finite number of small elements 
in each row and column by homogeneous configurations. 
Therefore with this trick one can generate the same 
number of coupled and uncoupled configurations. 

We can also generate new configurations with $N^{2}$ degrees 
of freedom starting from the homogeneous configurations 
and adding order $N$ large elements. For instance 
a unitary matrix with $N$ large elements and $N^{2}-N$ 
homogeneous elements. But with this trick we can 
also generate inhomogeneous configurations starting 
from inhomogeneous configurations: if we add 
order $N$ large elements to a matrix with order $N$ large 
elements the result will be a matrix with order $N$ large 
elements.

We can generate new configurations with order $N^{2}$ 
degrees of freedom from homogeneous configurations changing 
a complete row or column of homogeneous elements by 
small elements. If we apply this trick to an inhomogeneous 
configuration then the new configuration will have $N$ large 
elements and $N$ homogeneous elements and these will 
be coupled to the large elements. But the number of 
degrees of freedom will be also order $N$.                   

All the configurations generated from homogeneous 
and inhomogeneous configurations with combinations of the above 
tricks exhaust the set of unitary matrices. Therefore 
for each configuration with $N^{2}$ degrees of freedom 
generate from some homogeneous configurations we can 
generate a configuration with $N$ degrees of freedom 
from an inhomogeneous configuration. Hence 
the partition function in the large $N$ limit 
can be split into two terms:
\be
Z = A_{1} Z_{uncoupled} + Z_{coupled}
\ee
where $Z_{uncoupled}$ is the partition function restricted 
over the set of unitary matrices with only order $N$ nonzero 
elements. Actually 
\be
Z_{uncoupled} = N_{deg} \exp \{ - N Tr ( U^{\dagger} \Lambda U A \}
\ee
where $U$ is a solution of (\ref{eq:41}) and $N_{deg}$ is the 
number of solutions of (\ref{eq:41}). The partition 
function $Z_{coupled}$ is bounded by
\be
Z_{coupled} \le Vol \exp \{ - N  Tr ( U^{\dagger}_{S} \Lambda U_{S} A \}
\ee
where $U_{S}$ is the absolute minimum in the 
subset of unitary matrices with $N^{2}$ degrees of freedom. 
And $Vol$ is of the same order as $A_{1}$ in the large $N$ limit.

Hence if the absolute minimum of the 
potential which is given by (\ref{eq:41}) belongs to the 
subset of configuration with $N$ degrees of freedom then the partition 
function $Z_{uncoupled}$ gives the 
leading contribution to the partition function 
in the large $N$ limit. 

Let us study now the solutions of (\ref{eq:41}).
There are several cases: 
If the spectrum of the matrix $A$ is non degenerate, 
then the matrix $ U^{\dagger} \Lambda U $ must be a 
diagonal matrix because only a diagonal matrix can 
commute with a diagonal nondegenerate matrix. 
And if the spectrum of $\Lambda$ is non degenerate, 
then the identity matrix is the only 
solutions of (\ref{eq:41}). In this case 
the value of the matrix potential in (\ref{eq:36})
is $ Tr ( A \Lambda ) $. If the spectrum of $\Lambda$ 
is degenerate, then there are an infinite number of 
unitary matrices which 
satisfy $ U^{\dagger} \Lambda U = \Lambda $: all the 
unitary matrices corresponding to change of basis that 
leave invariant the subspace of eigenvectors of $\Phi$ 
with the same eigenvalue. These unitary matrices are 
block diagonal, the dimension of each block is given 
by the dimension of each subspace. Then 
if these dimensions are order $N$ then there are 
solutions of (\ref{eq:41}) that do not belong 
to the subset of inhomogeneous configurations 
and the reduction of the number of degrees of freedom 
does not take place. But this does not hold
because the Van der Monde determinant, which 
arise in the change of variables (\ref{eq:23}), 
forbids configurations where the matrix $\Lambda$
is degenerate. In the large $N$ limit the degeneration 
of $\Lambda$ cannot be order $N$ and the solutions 
of (\ref{eq:41}) belong to the 
subset of inhomogeneous configurations.    

If the spectrum of $A$ is degenerate, then  
the matrix $U^{\dagger} \Lambda U $ must be  
block diagonal. For instance if  
\bea
A_{ii} & = & a_{1} \ \ i = 1 , 
\cdots , n_{1} \nonumber \\
A_{ii} & = & a_{2} \ \ i = n_{1} , 
\cdots , n_{2} \nonumber \\
       & . & \nonumber \\
       & . & \nonumber \\
A_{ii} & = & a_{j} \ \ i = n_{j-1} , 
\cdots , N 
\label{eq:42}
\eea
then $U^{\dagger} \Lambda U $ is given by $j$ 
boxes in the diagonal of 
dimensions given by the set of numbers $n_{k}$. 
In this case the matrix $U$ is also 
block diagonal and the dimension
of each block depends on both 
matrices $A$ and $\Lambda$. If the degeneration 
of $A$ is order $N$ then there are solutions
of (\ref{eq:41}) which do not belong to the 
subset of inhomogeneous configurations, 
but we can chose the external matrix $A$ to be 
nondegenerate.  

In the large $N$ limit the eigenvalue configurations 
are given by continuum functions $a(x)$ 
and $\Lambda(x)$. Therefore near a given 
eigenvalue there are an infinite number 
of eigenvalues, but this number is order less than $N$. 
Therefore the solutions of (\ref{eq:41}) are unitary  
matrices which belong to the subset of inhomogeneous 
configurations in the large $N$ limit and the 
leading contribution to the partition function is given 
by $Z_{I}$.  

All the solutions of (\ref{eq:41}) give the same 
contribution to the partition function.
Actually the value of the 
matrix potential for solutions of (\ref{eq:41})
is always $Tr ( A \Lambda )$.
And one expect that the number of 
solutions must be a continuum functional of 
the eigenvalue configurations. 
Hence, in the large $N$ limit the partition 
function is 
\be 
N_{S} [ \lambda(x) , a(x) ] \exp \{ Tr ( \Lambda A ) \} 
\label{eq:43}
\ee
where $N_{S}$ is the number of solutions 
of (\ref{eq:41}). It is not possible to 
find the functional $N_{S}$ from the equation 
(\ref{eq:41}). But the integration in (\ref{eq:12}) 
can be performed exactly for finite $N$\cite{Itz}. 
Therefore, from the exact solution is possible to 
extract the functional $N_{S}$ in the large $N$ limit.

Let us remark a few things 
about the equation (\ref{eq:41}). In the application 
of one hermitian matrix models to condensed matter, 
mesoscopic and nuclear physics the matrix is 
identify with a Hamiltonian and the partition 
function is a sum over Hamiltonians which satisfy 
some symmetry\cite{Mat}. For instance matrix models 
with real matrices are related to models with 
time-reversal and spin-rotation symmetries, complex 
matrices are related to models with 
time-reversal symmetry broken by a magnetic field 
or magnetic impurities, and quaternionic 
matrices are related to models with the spin-rotation 
symmetry broken by strong spin-orbit scattering 
for instance. 
In all this cases the correlation between the 
eigenvalues $\{ \lambda_{i} \}$ are  given by 
\be
\prod_{i \ne j } 
( \lambda_{i} - \lambda_{j} )^{\beta}
\label{eq:index}
\ee
where $\beta$ can be $1$, $2$ or $4$ 
depending on the symmetry of the model. 
Actually $\beta$ is the number of degrees of 
freedom in the matrix elements: $1$ for real 
matrices, $2$ for complex matrices and $4$ for 
quaternionic matrices. But it is well known 
that if one introduce into the matrix potential 
a coupling with an external matrix, as 
in (\ref{eq:10}), the exponent in (\ref{eq:index}) 
changes. But let us observe that, 
in the large $N$ limit, the coupling with an 
external matrix restricts the sum over all 
hermitian matrices to the sum over hermitian matrices 
that commute with the external matrix $A$ (\ref{eq:41}). 
Therefore the change in the index $\beta$ is still related with 
a change in the  symmetry of the set of matrices over which 
the path integral is defined.    

Let us perform the calculation of 
\be 
Z = \int D \Phi \exp \left 
\{ - N Tr ( \Phi A \Phi A + V(\Phi) ) 
\right \}
\ee  
Now the integral over the unitary 
matrices $U$ is given by:
\be 
\tilde{Z} = \int D U \exp \left \{ - N Tr ( 
U^{\dagger} \Lambda U A U^{\dagger} 
\Lambda U A ) \right \}
\label{eq:44}
\ee

One can prove that the classical 
configurations of the 
matrix potential are given by
\be
 ( U^{\dagger} \Lambda U A ) ^ 2 = 
( A U^{\dagger} \Lambda U ) ^ 2.
\label{eq:45}
\ee

The solutions of (\ref{eq:41}) are also 
solutions of the above equations. 
But there are other solutions. 
For instance, the configurations which 
satisfy 
\be 
U^{\dagger} \Lambda U A = 
- A U^{\dagger} \Lambda U
\label{eq:46}
\ee
are also solutions of the new saddle-point 
equation. In this case the number of solutions 
depends on the diagonal matrix $A$ and $\Lambda$. 
For instance, if $A$ is the identity matrix 
there are not solutions of (\ref{eq:46}). 
Actually, there are solutions of (\ref{eq:46}) 
only if:
\be
a_{ii} = - a_{jj}
\label{eq:47}
\ee
for some $i$ and $j$. Therefore the form 
of the matrix $A$ must be:
\be
\left (
\begin{array}{cccccccc}
A_{1}&&&&&&&\\
&-A_{1}&&&&&&\\  
&&A_{2}&&&&&\\
&&&-A_{2}&&&&\\
&&&&\cdot&&&\\
&&&&&\cdot&&\\
&&&&&&\cdot&\\
&&&&&&&D\\
\end{array}
\right )
\label{eq:48}
\ee
where $A_{i}$ are matrices proportional 
to the identity matrix and 
$D$ is a matrix which does not 
verify (\ref{eq:47}). Then the form of 
the matrix $U^{\dagger} \Lambda U$ must be
\be
\left (
\begin{array}{cccccccc}
0&B_{1}&&&&&&\\
B^{\dagger}_{1}&0&&&&&&\\  
&&0&B_{2}&&&&\\
&&B^{\dagger}_{2}&0&&&&\\
&&&&\cdot&&&\\
&&&&&\cdot&&\\
&&&&&&\cdot&\\
&&&&&&&0\\
\end{array}
\right )
\label{eq:49}
\ee
where $B_{i}$ is a complex matrix, 
and the square root of the eigenvalues 
of the hermitian matrices 
$B^{\dagger}_{i} B_{i}$ are the eigenvalues of $\Lambda$.
The value of the matrix potential for these 
configurations is 
\be
- 2 \sum_{i} a_{i}^{2} Tr 
( B^{\dagger}_{i} B_{i}) \ne Tr ( A^{2} \Lambda^{2} ),
\ee
where $A_{i} = a_{i} I$. Whereas 
for the same matrix $A$ and $\Lambda$, 
the value of the action for the solution of (\ref{eq:41}) 
is 
\be
2 \sum_{i} a_{i}
Tr (B^{\dagger}_{i} B_{i} ) = Tr ( A^{2} \Lambda^{2} ). 
\ee
Therefore, solutions of
\bea
U^{\dagger} \Lambda U A & = 
& A U^{\dagger} \Lambda U \nonumber \\
U^{\dagger} \Lambda U A & = 
& - A U^{\dagger} \Lambda U 
\label{eq:50}
\eea
are saddle-point configurations of 
the partition function (\ref{eq:12}).
But their contributions to the 
partition function are different. Therefore,
we can think the two set of solutions as 
different vacua of the model. There are other
saddle-point configurations: 
configurations for which the matrix 
$U^{\dagger} \Lambda U A $ has 
two blocks in the diagonal, in such a way that 
one of the block satisfy the 
commutation conditions and the other
the anticommutation conditions. 
These sets of solutions give different
contribution to the 
integration over the $U$ matrix. 

This is an interesting difference between 
the linear coupling to an external matrix 
and the gaussian coupling: the former has 
only one vacuum whereas in the gaussian coupling 
there are several vacua of the saddle point 
equation corresponding to the integration of 
the unitary matrix. But in all this cases if 
the degeneration of the external matrix 
is order one in the large $N$ limit then 
the solutions of saddle point equation are 
given by inhomogeneous configurations 
and the reduction of degrees of freedom holds. 
If the external matrix $A$ does not verify (\ref{eq:47}) 
then the only solutions are given by solutions of    
\be
U^{\dagger} \Lambda U A = A U^{\dagger} \Lambda U 
\label{eq:51}
\ee
In this case the eigenvalues must be all of the same 
sign because if for some $i$ and $j$, the corresponding 
eigenvalues verify $\lambda_{i} = - \lambda_{j}$ then 
there are solutions different from (\ref{eq:51)}. This 
is because the symmetry between $A$ and $\Lambda$ in the  
equations.   
For this vacuum the degeneration 
factor $N_{S}[\Lambda,A]$ must the 
same as in the linear case.  
Therefore the partition function (\ref{eq:12}), 
in the large $N$ limit, becomes:
\be
\tilde{Z} \propto \frac{1}{\Delta(\Lambda)} 
\exp \{ - N Tr (A^{2} \Lambda^{2} )\}
\label{eq:52}
\ee

Even though this model can be used to define 
pure Quantum Gravity with extrinsic curvature 
its solution is not very interesting from a physical 
point of view\cite{Kaz, Bow}.

{\bf 4. Conclusions}

The main conclusion of this paper is that 
for matrix models with $N^{2}$ degrees of 
freedom for finite $N$ there is a reduction 
of the number of degrees of freedom in 
the large $N$ limit, actually the number of 
degrees of freedom becomes order $N$. 
This reduction is the origin of the 
factorization of the vacuum expectation 
values of products of observables in the 
path integral approach in the matrix models 
studied in this paper. 

The reduction showed in this paper is different 
from the well known Eguchi-Kaway reduction\cite{EK}. 
The EK reduction is a reduction of the external 
space: the model reduces to a zero dimensional 
model; while in the reduction studied in this 
paper the reduction takes place in the internal 
space of degrees of freedom: actually the matrix 
models studied in this paper are zero dimensional. 

It seems that this reduction is more fundamental 
than the EK reduction. This reduction is the origin 
of the factorization of product of observables and the 
factorization is the origin of the EK reduction.     

This result holds only if at least order $N^{2}$ 
degrees of freedom can be thought as the elements 
of some unitary matrices and the absolute minimum 
of the matrix potential, as a function of this 
unitary matrices, is given by a configuration with 
only order $N$ degrees of freedom. The first condition 
is accomplished by every matrix model and also by 
gauge theories in the lattice because every hermitian 
or unitary matrix 
can be split into the diagonal matrix of its 
eigenvalues and a unitary 
matrix:$ \Phi = U^{\dagger} \Lambda U$. 
The second condition seems to depend on 
the matrix potential. For instance, 
in two unitary matrix models with 
matrix potential given 
by 
\be
Tr V (U_{1}) + 
Tr V ( U_{2}) + Tr (U_{1} U_{2} ) + h.c., 
\ee
there are saddle point configurations which 
do not verify the second condition. 
But if one perform the following 
change of variables
\be
U_{i} \rightarrow 
\Omega^{\dagger}_{i} \theta_{i} \Omega_{i},
\label{eq:c1}
\ee
where $\theta_{i}$ is the diagonal matrix 
of the eigenvalues of $U_{i}$ and $\Omega$ 
is a unitary matrix; then the saddle point 
configurations corresponding to the matrix 
variable $\Omega$ verify the second condition. 
Actually this problem is analogous to the 
matrix models studied in this paper. Therefore
if for any given matrix model or gauge 
theory in the lattice is possible to find 
a change of variables, as for 
instance (\ref{eq:c1}), such that order $N^{2}$ 
degrees of freedom are the elements of some unitary 
matrices, for instance $\Omega$ in (\ref{eq:c1}), 
and perhaps there are other $N$ degrees of freedom, 
for instance the diagonal matrix $\theta$ 
in (\ref{eq:c1}), and the saddle point 
configurations of the potentials as functions 
of the former unitary matrices are given by $N$ 
degrees of freedom only, then the reduction of 
the number of degrees of freedom holds in the 
large $N$ limit. 

This reduction must be a general result 
of the large $N$ limit because the factorization 
of the expectation values of products of 
observables is a general result and is difficult 
to understand how the product of observables depending 
on $N^{2}$ degrees of freedom can factorize if 
fluctuation to all orders enter into the path integral. 
Therefore we can conjecture that this reduction in 
the internal space must be a general behaviour  
of the large $N$ limit in the path integral approach.

\vfill

\pagebreak

\end{document}